\documentclass[useAMS,usenatbib,usegraphicx]{mn2e}

\newcommand{\cl}{C_{\ell}}
\newcommand{\alm}{a_{\ell m}}
\newcommand{\wl}{W_{\ell}}
\newcommand{\chieff}{\chi^2_{n_{eff}}}
\newcommand{\neff}{n_{eff}}

\newcommand{\like}{\mathcal{L}}

\newcommand{\vy}{ \bf{ \hat{y} } }
				
\newcommand{\valp}{\bf{ \hat{\alpha}} }

\newcommand{\mx}{\bf { \hat{x}}}
\newcommand{\mM}{\bf { \hat{M}}}
\newcommand{\mV}{\bf { \hat{V}} \it{}}
\newcommand{\mW}{\bf{ \hat{W}} \it{}}

\title[A Modified $\chi^2$-Test for CMB Analyses]{A Modified 
$\chi^2$-Test for CMB Analyses}
\author[J.A. Rubi\~no-Mart\'{\i}n and J. Betancort-Rijo]{J.A. 
Rubi\~no-Mart\'{\i}n$^{1}$\thanks{Present address: Max-Planck-Institut
f\"ur Astrophysik, Karl-Schwarzschild Str. 1, D-85748 Garching,
Germany.}\thanks{E-mail:jalberto@ll.iac.es, jalberto@MPA-Garching.MPG.DE} 
and J. Betancort-Rijo$^{1,2}$\\
$^{1}$Instituto de Astrofisica de Canarias, C/ Via Lactea, s/n, 38200
La Laguna, Tenerife, Spain\\
$^{2}$Departamento de Astrofisica, Universidad de La Laguna, 38200 La
Laguna, Tenerife, Spain}

\begin{document}

\date{}

\pagerange{\pageref{firstpage}--\pageref{lastpage}} \pubyear{2003}

\maketitle

\label{firstpage}

\begin{abstract}
We present a new general procedure for determining
a given set of quantities. 
To this end, we define certain statistic,
that we call 'modified $\chi^2$' ($\chi^2_M$),
because of its similarity
with the standard $\chi^2$.
The terms of this $\chi^2_M$ are made up of the 
fluctuations of an unbiased estimator of some statistical 
quantities, and certain weights. 
Only the diagonal terms of the covariance matrix
explicitly appear in our statistic, while the full 
covariance matrix (and not its inverse) 
is implicitly included in the calculation of the weights. 
Choosing these weights 
we may obtain, through minimising the $\chi^2_M$, the 
estimator that provides the minimum RMS, 
either for those quantities or for the
parameters on which these quantities depend.
In this paper, we describe our method in the context
of Cosmic Microwave Background experiments, in order
to obtain either the statistical properties of the maps, or 
the cosmological parameters.
The test here is constructed out of some
estimator of the two-point correlation function at 
different angles. 
For the problem of one parameter estimation, we show that
our method has the same power as 
the maximum likelihood method.
We have also applied this method to Monte Carlo 
simulations of the COBE-DMR 
data, as well as to the actual 4-year data,
obtaining consistent results with previous analyses.
We also provide a very good analytical approximation to 
the distribution function of our statistic, which could also
be useful in other contexts. 
\end{abstract}


\section{Introduction}

The study of the Cosmic Microwave Background (CMB) anisotropies 
is providing strong constraints on theories of structure formation.
These theories are statistical in essence, so the extraction of the 
information must be done in a statistical way. In particular, the 
standard method for analysing a CMB experiment is the maximum 
likelihood estimator (ML). The procedure is straightforward: 
maximise the probability of the parameters of the model 
given the data, P(parameters$|$data), 
over the allowed parameter space. Usually, we take the prior 
probability for the parameters to be constant, so this is 
equivalent to maximising the
likelihood, P(data$|$parameters), via the Bayes' theorem.

The ML method has been widely applied in CMB analyses, 
for power spectrum or parameters estimation,  
\citep{davies87,gorski94,hinshaw96a}.
When computing the likelihood in these problems,
we have to deal with the inversion of the 
covariance matrix of the data,
which usually involves $O(N^3)$ operations, 
being $N$ the number of pixels of the map.
The increasing size of the datasets 
makes this method computationally costfull for 
new experiments, so other methods have been investigated
in the last few years to confront the problem. 
There have been several proposals on this matter.
Pioneering work on the problem of power spectrum estimation 
\citep{hp73,peebles73},
based on an evaluation of the $\alm$'s coefficients 
of the multipole expansion of the observed map in the spherical harmonics  
basis, 
have been applied to COBE data \citep{wright96}.
Quadratic estimators have been proposed by several authors 
\citep{tegmark97, bond98} as statistics that give the same 
parameters that maximise the likelihood, but requiring less computational
work.

Nevertheless, alternative statistical methods are required
in the field to extract the cosmological information from future 
CMB experiments (as PLANCK) where the number of data points will be very
large (see, e.g. \citet{borrill99} for an estimation of the scaling
of the computing time with the dataset size).

Here, we propose a new statistical method to analyse a CMB map.
In order to illustrate it, we will use 
the two-point correlation function (CF).
We first replace the likelihood of the full map by the likelihood of the
fluctuations of an estimator of the CF. Then, we derive the cosmological
parameters from it in an efficient manner. 
If we assume gaussianity for the primordial CMB fluctuations, the CF
completely characterises the statistical properties of 
the field. In this line, it has been suggested \citep{bashinsky01}
that it can be used to obtain the power spectrum 
or for parameter estimation, because it encodes all the relevant 
information for that purpose.
This approach of considering the CF in CMB analyses  
has been recently used by other authors 
\citep{szapudi01,atrio01,szapudi01b}
to estimate the power spectrum. 
They obtain the CF using different estimators,  
and integrate it, projecting over the Legendre polynomials,
to obtain the $\cl$'s. The advantage of this 
estimator is that it only 
needs at the most $O(N^2)$ operations 
to be computed, and not $O(N^3)$, as is 
required for the likelihood. 

We construct a modified version of the standard $\chi^2$ test, 
using the CF evaluated at a 
certain set of points. The estimate of the parameters of the model is given
by the minimum of this statistic, as in the standard analysis. 
We give a very good approximation to the 
distribution function of this 
modified $\chi^2$, so the confidence limits can be obtained 
without using simulations, by integration below that curve, as for the ML. 
We show that, in several problems, choosing a large enough 
set of points to evaluate the CF, our method has 
the same power as the maximum likelihood, while
being two different methods.

\section{The modified $\chi^2$-test}
In this section we will introduce the test, 
using for this purpose the two-point CF. 
Nevertheless, all the procedure described below
can be applied to any other estimator.

For a certain map of the CMB anisotropies, $X = \{ x_1, ... , x_N \}$ 
with $N$ pixels, and errors $\sigma = \{ \sigma_1, ... , \sigma_N \}$, 
we can estimate the CF, $C(\theta)$, in 
a set of $n$ angular distances, 
$\{ \theta_k \}_{k=1}^{n}$.
In this work we have used the following estimator, 

\begin{equation}
E[ C(\theta_k) ] = \frac{ \sum_{ i,j \in \{ k \} }  \; 
x_i x_j}{ \sum_{ i,j \in \{ k \} } 1 }
\label{est_cf}
\end{equation}

\noindent
where $\{ k \}$ stands for the set of all pixel 
pair $(i,j)$ such that their 
angular distance is $\theta_k$, 
but our proposal and techniques can be applied to other 
estimators for the CF (see, for example, \cite{atrio01}, 
or \cite{szapudi01b}).
Hereafter, we will write the estimate of a certain 
parameter as $E[...]$.
If we have a map with zero mean and no noise, 
eq. (\ref{est_cf}) is a unbiased estimator of the theoretical two-point 
CF, which for a experiment with a symmetric beam is given by 

\begin{equation} 
C(\theta_k) = \frac{1}{4\pi} \sum_{\ell = 2}^{+\infty} (2\ell +1) \cl \wl 
P_\ell ( \cos \theta_k ) 
\label{cf_th}
\end{equation} 
where $\wl$ stands for the window function of 
the experiment. We have explicitly removed the 
dipole contribution in the previous equation, 
because these coefficients are dominated by
the kinematic dipole in a real map.
If the CMB signal is gaussian, the power spectrum 
(or its Fourier transform, the CF), 
encodes all the information about the model. 
So, under this assumption, we can parameterise a model through 
the $\cl$'s themselves, or through the cosmological parameters, by writing 
$\cl = \cl( n, \Omega, \Omega_b, \Omega_{\Lambda}, H_0, ...)$. 
In general, we will write 
$C(\theta_k|\hat{M})$, being $\hat{M}$ the parameters of the model. 

As an example of the general procedure that we propose in this paper, 
we consider in detail the estimator derived from the following statistic:

\begin{equation}
\chi^2_M (\hat{M}) = \sum_{k=1}^{n} P_k
\frac{ ( E[ C(\theta_k) ] - C(\theta_k|\hat{M}) - C_N(\theta_k) )^2 }
{\sigma^2( C(\theta_k) ) }
\label{chi2}
\end{equation}
 
\noindent
where $P_k$'s are certain weights to be defined below, and 
$\sigma^2( C(\theta_k) )$ stands for the variance of the 
estimator $E[ C(\theta_k) ]$. $C_N(\theta_k)$ represents the discrete 
CF of the noise. If we have an experiment with uncorrelated noise,
this function takes the form

\begin{equation}
C_N(\theta_k) = \Bigg \{
\begin{array}{lr} 
\frac{ \sum_i \sigma^2_i }{N},  & \theta_k = 0 \\
0,  & \theta_k \ne 0 
\end{array}
\label{cf_noise}
\end{equation}

This method is a modification of the standard form of a 
$\chi^2$-test, for the case when 
the error of each of the estimates 
entering (\ref{chi2}) are independent 
and gaussianly distributed (hereafter, we mean by standard 
$\chi^2$-test the case when $P_k = 1$, $\forall k$, 
and all the terms of 
the sum in (\ref{chi2}) are independent).
In the present case, the $E[ C(\theta_i) ]$ quantities 
follow very closely a gaussian, but are correlated. 
For this reason, we have introduced some weights ($P_k$), that
will be determined by minimising the dispersion of the estimator
derived from equation (\ref{chi2}), as we will see in 
the next section.
Those quantities will account for the different degree of 
correlation between terms, and 
in a general problem, they 
will be a function $P_k = P_k(C')$, where 
$C'_{ij}$ stands for the correlation matrix 
between the errors of the estimates 
of $C(\theta_i|\hat{M})$ and $C(\theta_j|\hat{M})$, i.e.
\begin{eqnarray}
\nonumber
C'_{ij} = C'_{ij}(\hat{M})= 
< \Bigg( E[ C(\theta_i) ] - C(\theta_i|\hat{M}) 
- C_N(\theta_i) \Bigg) \times \\
\times \Bigg( E[ C(\theta_j) ] - C(\theta_j|\hat{M}) 
- C_N(\theta_j) \Bigg) > 
\label{cprima}
\end{eqnarray}

\noindent
so the variance $\sigma^2( C(\theta_k) )$ is related to the 
$C'$-matrix by 
\begin{equation}
\sigma^2( C(\theta_i) ) = C'_{ii}
\end{equation}

\noindent
The brackets $<...>$ represent an average 
over an ensemble of Universes, i.e.,
an average over realizations for one fixed CMB model.

In principle, our construction seems to 
miss information about the correlations when
compared to the usual $\chi^2$ procedure 
to analyse correlated datasets, because 
in equation (\ref{chi2}) only the diagonal 
terms of the covariance matrix ($C'$) are 
explicitly shown. 
Nevertheless, as we will see in the next section, 
the $P_k$ weights depend on the full covariance matrix  
and not on its inverse, so all the correlations
implicitly enter in that expression. 

We will not make a detailed comparison between 
the usual $\chi^2$ method with uses
the full covariance matrix C' (we will refer to this method as 
the ``usual $\chi^2$'') and our $\chi^2_M$. However, we will illustrate 
this with an example (see section 6.1). 
In addition, in Appendix A we present a  
brief comparison of some characteristics of both methods 
($\chi^2_M$ and the usual $\chi^2$) for the case of linear problems.
It is an interesting result that, for gaussian linear problems,
if we are estimating only one parameter, 
the estimates from both methods are exactly equal, 
while being different statistics (i.e. they will give
different probability contours). 
Hereafter, we will concentrate in our method, and its 
application to CMB problems.

It is worth to notice that our $\chi^2_M$ statistic is a different
approach to the ML, in the sense that it provides different estimates
and probability contours. However, in the case of CMB analyses, we will
see below that it has a similar power to the ML, 
but avoiding the problem of the inversion of the covariance matrix. 
There is however a minor sense in which our test may formally be
considered as an approximation to the ML. It is well-known that the ML
is an asymptotically efficient estimator for our problem.
Thus, in the limit of infinite size of the sample (or for those
problems where an efficient estimator exists, 
as in linear gaussian problems), 
then the ML is the only one statistic which renders the minimum
variance, and thus any other estimator may be regarded as approximate.

\subsection{Estimate of the method}

Once we have constructed the $\chi^2_M$ test, the estimate for 
this method is given by the set of values 
for the parameters that minimise eq. (\ref{chi2}), i.e. 
the solution to the set of equations 
$\partial \chi^2_M (\hat{M}) / \partial \hat{M} = 0$.
For a given problem, we proceed as follows. 
If we want to estimate a set of $p$ parameters, 
$\hat{M} = \{ \xi _1, ..., \xi _{p} \} $, we 
first compute the $C'$ matrix by assuming an initial value for 
those parameters, $\hat{M}_0$.
Using this matrix, we obtain our estimate by solving the following 
system of equations,

\begin{equation}
\frac{\partial \chi^2_M}{\partial \xi _i} = 0, \qquad i = 1,...,p
\label{ps_est}
\end{equation}

\noindent
When computing these derivatives with 
respect to $\hat{M}$, we neglect
the dependence of $C'$ on the parameters, 
which is equivalent to assuming
that $\sigma^2( C(\theta_k) )$ and $P_k$ are constants in 
the derivation.
This process is iterated until convergence.
The reason to keep the $C'$ matrix fixed in the derivation is that 
we want to have an unbiased estimator of the parameters.

The evaluation of the $C'$ matrix can be done by Monte Carlo simulations, 
but we also propose an analytical approach. It is possible 
to evaluate  equation (\ref{cprima}) using the 
quantities $<x_i x_j x_k x_l>$, for a multivariate-gaussian field, as 
is the case for the CMB (see Appendix B). 

In the particular case of power spectrum estimation, 
the set of parameters we have to determine are the 
$\cl$'s themselves, or the band powers in a certain 
number of multipole bands centred at multipoles 
$\{ \ell_1, ..., \ell_{p} \}$ 
(i.e., $\hat{M} = \{ \cl  \}_{\ell=\ell_1}^{\ell=\ell_{p}}$). 
As the theoretical CF (\ref{cf_th}) is linear in 
the $\cl$'s, we have that eq. (\ref{ps_est}) is a linear 
system of $p$ equations, and the solution can easily 
be found.

As an example, we present here the equations for the determination 
of the total power measured by a certain experiment. In this case, we only 
have one parameter, $\hat{M} = \{ \sigma_{sky}^2 \} $, defined as  

\begin{equation}
\sigma_{sky}^2 = C(\theta = 0^o) = (\Delta T / T)^2_{RMS} = 
\sum_\ell \frac{2\ell+1}{4\pi} \cl \wl 
\label{def_ssky}
\end{equation}

\noindent
which is essentially a normalisation of the spectrum.
From here, we define the function 
$f(\theta) \equiv C(\theta) / \sigma_{sky}^2$, which is independent
of $\sigma_{sky}^2$.
We can now obtain the analytic expression for the estimate of 
$\sigma_{sky}^2$ by minimising eq. (\ref{chi2}) with respect to 
$\sigma_{sky}^2$, which in this case takes the form

\begin{equation}
\chi^2_M ( \sigma_{sky}^2) = \sum_{k=1}^{n}
P_k \frac{ ( E[ C(\theta_k) ] - \sigma_{sky}^2 f(\theta_k) ) ^2}
{\sigma^2( C(\theta_k) ) }
\end{equation}

\noindent
For simplicity, we will not write the term of the noise CF, but
it can be easily included inside the true CF.
Inserting the previous expression in equation (\ref{ps_est}), we obtain, 
for fixed $P_k$, an equation for $\sigma_{sky}^2$. It must be noted
that due to the dependence of $\sigma^2( C(\theta_k) )$ on 
$\sigma_{sky}^2$, this equation is not exactly linear. We could
solve it iteratively, starting with certain fixed value for
$\sigma^2( C(\theta_k) )$. However, for all the values of these 
quantities within the current limits, a first iteration is enough, as 
we will see, so that the equation determining $\sigma_{sky}^2$ is
effectively linear, and therefore its solution is given by

\begin{equation}
E[\sigma_{sky}^2] = 
\Bigg( \sum_i \frac{ P_i f_i^2 }{ \sigma^2( C(\theta_i) )} \Bigg )^{-1} 
\sum_i P_i f_i \frac{ E[ C(\theta_i) ] }{ \sigma^2( C(\theta_i) ) } 
\label{E_ssky}
\end{equation}

\noindent
The $RMS$ of this estimator is given by

\begin{eqnarray}
\nonumber
RMS( E[\sigma_{sky}^2] ) = 
\Bigg( \sum_i \frac{ P_i f_i^2 }{ \sigma^2( C(\theta_i) )} \Bigg )^{-1}
\times \\ 
\qquad \times \Bigg[ \sum_i \frac{P_i^2 f_i^2}{\sigma^2( C(\theta_i) )} +
\sum_{i \neq j} \frac{ f_i f_j P_i P_j C'_{ij} }
{\sigma^2( C(\theta_i) ) \sigma^2( C(\theta_j) )} \Bigg]^{1/2}
\label{rms_ssky}
\end{eqnarray}

\noindent
where we have defined $f_i = f(\theta_i) $.
The expression within large parentheses 
in this equation 
is the $RMS$ of the second sum in equation (\ref{E_ssky}).
The first sum within the parentheses correspond
to the quadratic addition of the contributions
of each term in (\ref{E_ssky}), which is present 
even when the random variables $E[ C(\theta_i) ]$
are independently distributed. The second sum is
due to the correlations between any pair of 
these variables.  
It should be noted that equation (\ref{rms_ssky}) has been
obtained assuming that the quantities $E[ C(\theta_i) ]$ follow 
a multivariate gaussian distribution.
This is a good approximation if there are
enough pixel pairs entering in the sum in (\ref{est_cf})
(see, for example, \cite{hinshaw96b}, for the CF of the COBE data). 
Similar calculations for the 
standard $\chi^2$ and the likelihood function can be found 
in \cite{b93} (hereafter, B93). The matricial expression of the 
estimate and the RMS for a general linear problem are shown in
Appendix A.

\section{The $P_k$ quantities for a given problem }

The $P_k$'s weights in equation (\ref{chi2}) are introduced 
in order to take into account the different degree of 
correlation of the terms of the sum.
Their expression can be obtained once we define 
exactly what we are interested in. 
For example, one common criteria for one parameter 
estimation is to use the estimator which has the minimum $RMS$.
We will consider this criteria here.

For the problem of one parameter estimation described in
the previous section, once we have the 
analytic expression for the $RMS$, and an initial guess
for the $C'$ matrix, we can obtain the optimum 
set of $P_k$'s using the ``minimum $RMS$ criteria''.
We minimise eq. (\ref{rms_ssky}) with respect to 
the $P_k$'s quantities.
We obtain that the $P_k$'s quantities are given 
by the solution to the implicit set of equations

\[ 
\nonumber
P_k f_k \sum_i \frac{P_i f_i^2}{\sigma^2( C(\theta_i) )} + \frac{1}{2}
\Bigg( \sum_{i \ne k} \frac{f_i P_i C'_{ik}}{\sigma^2( C(\theta_i) )} \Bigg) 
\Bigg( \sum_j \frac{P_j f_j^2}{\sigma^2( C(\theta_j) )} \Bigg) - 
\]
\begin{equation}
- f_k \sum_i \frac{f_i^2 P_i^2}{\sigma^2( C(\theta_i) )} - 
f_k \sum_{i \ne j} \frac{f_i f_j P_i P_j C'_{ij}} 
{\sigma^2( C(\theta_i) ) \sigma^2( C(\theta_i) )} = 0, \qquad  k=1,...,n 
\label{P_k_ssky}
\end{equation}
which can be solved numerically, using a Newton-Raphson scheme for nonlinear
systems of equations.
It should be noted that in the case when 
$C'_{ij} = 0$, $i \ne j$, equation (\ref{P_k_ssky}) 
has the trivial solution $P_k = 1$, as we expected for
the standard case without correlations.
The estimates obtained with these $P_k$ give us a better guess for
$C'$, that could be used in equation (\ref{P_k_ssky}) to obtain 
more appropriate values of the $P_k$. However, in practice, 
we have checked that for all the cases that we consider in this paper, 
this iteration is not necessary, since over the a priori uncertainty region
of the parameter, the variation of the $P_k$ is negligible.

The previous expression, derived for the problem of total power
estimation, can also be applied to any problem of one parameter
estimation, as follows. Let $M$ be the parameter we are interested in.
If we expand the CF in a Taylor series around 
an initial guess, $M=M_0$, we obtain, up to first order, 

\[
\Delta C(\theta_k | M) = C(\theta_k | M) - C(\theta_k | M_0) =
\]
\begin{equation}
 = \frac{ \partial C(\theta_k | M)}{\partial M} 
\Bigg |_{M_0} \Delta M  + O(\Delta M^2), \qquad k=1,...,n
\end{equation}
with $\Delta M = M - M_0$, 
so we can use equation (\ref{P_k_ssky}), with 

\begin{equation}
f_k = \frac{ \partial C(\theta_k | M)}{\partial M}\Bigg |_{M_0},
\qquad k = 1,... n
\end{equation}

\indent
If we use an initial guess close to the real value, this linear 
approximation will give good results.
The $f_k$'s can be obtained numerically for each problem.

When we deal with a problem of several 
parameters estimation, it is not well defined what 
has to be minimised.
A reasonable criteria for these problems, if we want to estimate
the set $\hat{M} = \{ M_1,...,M_p\}$,
is to minimise $\prod_i RMS( E[M_i] )$, 
where we define $RMS( E[M_i])$ as the 
$RMS$ for each individual parameter.
Linearising around our initial guess, $\hat{M}_0$, we can 
derive a simple expression for the $RMS$ of each parameter.
In this case, we have that 

\begin{equation}
\Delta C(\theta_k | \hat{M}) = \sum_{i=1}^{p} f_{k,i}
\Delta M_i  + O(\Delta M^2), \qquad k = 1,... n
\end{equation}
where we define 

\begin{equation}
f_{k,i} = \frac{ \partial C(\theta_k | \hat{M})}{\partial M_i} 
\Bigg |_{\hat{M}_0}, \qquad i=1,...,p
\end{equation}

\noindent
The estimate of the parameters is given
by the solution to the linear system of equations 
$\partial \chi^2_M / \partial M_i = 0$, where we have again neglected
the dependence of $\sigma^2(C(\theta_k))$ on the parameters.
The general expression of the covariance matrix of the parameters
is shown, for a general linear problem, 
in Appendix A. If we expand those matrices, we obtain that the
general form of this estimate, for our problem, is given by

\begin{equation}
E[\Delta M_i] = \frac{1}{R} \sum_{k} J_{k,i} P_k 
\frac{ E[ C(\theta_k)] - C(\theta_k,\hat{M}_0)}
{\sigma^2(C(\theta_k)) }, 
\qquad i=1,...,p
\label{est_M_k}
\end{equation}
where $E[\Delta M_i] = E[M_i] - (M_0)_i$, and 
$R$ and $J$ are numbers obtained from the $f_{k,i}$'s.
From (\ref{est_M_k}) we can infer 
the general expression
for the $RMS$ in the case of several parameters, obtaining

\begin{equation}
RMS( E[M_i] ) = 
\frac{1}{R} \Bigg[ \sum_{k,j} J_{k,i} J_{j,i} P_k P_j
\frac{ C'_{kj} }{\sigma^2(C(\theta_k))  \sigma^2(C(\theta_j)) }
\Bigg]^{1/2}
\label{rms_several}
\end{equation}
where $i=1,...,p$. 
Using the previous equation, we can obtain the $P_k$ quantities
for any problem, just minimising the product 
$\prod_i RMS( E[ M_i ] )$
numerically. Summarising, we will have a
different expression for the $P_k$'s
for each particular problem. 
An application of these equations 
for the problem of two parameters estimation 
can be found in Section 6.1.

\section{Distribution function for the modified $\chi^2$ }

The $\chi^2_M$ proposed in eq. (\ref{chi2}) 
corresponds to a sum of quantities 
which are not independent. If we had set $P_i = 1$, we would have the 
standard $\chi^2$ statistic, but still with correlations among the terms.
Therefore, its distribution function will not be the standard one. 

The formal expression for the 
distribution of a $\chi^2$ constructed from variables which are
distributed following a multivariate gaussian distribution is 
given in Appendix C.
This distribution, when applied to CMB analyses, was 
studied in B93. There, they proposed 
that the distribution function for the 
statistic (\ref{chi2}) is given by 
an standard (rescaled) $\chi^2$ function, 
but with an effective number of degrees of freedom. 
This proposal is not exact (see also Appendix C), 
but it turns out to be a very good 
approximation for the true distribution function, 
as we shall see in the following section. 
In a general case, the error 
in the distribution function using our approximation 
will be a few percent.

The basis of the approximation is to assume that correlations
only reduce the degrees of freedom, but do not change the shape of the 
distribution \footnote{This idea has been used recently by other 
authors: \cite{wandelt00,hivon01}.}. 
Quantifying this argument, there exist a certain constant, $A$,
which makes the statistic 

\begin{equation}
U = A \chi^2_M
\end{equation}

\noindent
to be distributed as an ordinary
$\chi^2$, with an effective number of degrees of freedom, $n_{eff}$.
This number, and the constant $A$, are obtained just by imposing
the new distribution to have the mean and the variance of a standard 
$\chi^2$, i.e., 

\begin{equation}
 < U > = n_{eff}, 
\end{equation}
\begin{equation}
MS(U) \equiv < U^2 > - < U >^2 = 2 n_{eff}
\end{equation}

\noindent
where $MS$ means mean square. Summarising, our proposal
is that once the first and second moments 
are fixed, the whole distribution will follow a $\chi^2$
very closely. For our problem, we obtain 

\begin{equation}
A = \frac{2 <\chi^2_M> }{ MS(\chi^2_M) } = 
\frac{ \sum_{i=1}^{n} P_i }
{ \sum_{i=1}^{n} P_i^2  +  \sum_{i \neq j} P_i P_j   
\frac{ C_{ij}^{'2} }{ \sigma^2( C(\theta_i) ) \sigma^2( C(\theta_j) ) }  }
\label{A}
\end{equation}

\begin{equation}
n_{eff} = \frac{2 <\chi^2_M>^2 }{ MS(\chi^2_M) } = 
A \sum_{i=1}^{n} P_i
\label{neff}
\end{equation}

\noindent
In this calculation, we needed to compute the $MS$ 
of eq. (\ref{chi2}). This result is obtained using 
the fact that the data points follow a multivariate gaussian
distribution, and it can be found in B93 (see also Appendix B for
similar calculations).

In general, $n_{eff}$ is a real number, so we have 
to consider the analytic 
extension of a standard $\chi^2$ distribution 
(which is known as the Gamma distribution function, 
see for example \cite{kendall}),

\[
dF(\chieff,\neff) = g(\chi^2_{n_{eff}}, n_{eff}) d\chieff = 
\]
\begin{equation}
\frac{1}{2^{\neff} \Gamma (\frac{\neff}{2}) } 
\exp ( -\frac{\chieff}{2} ) 
(\chieff)^{\frac{\neff}{2} - 1} d\chieff
\label{dFchi2}
\end{equation}
just by replacing the factorial with the gamma function ($\Gamma$).
In (\ref{dFchi2}), $dF$ is the probability of finding a value for 
$U$ between $\chieff$ and $\chieff + d\chieff$, 
and $g$ stands for the probability density function.
Once we know the distribution function, the confidence limits are given by
integration bellow this curve. 
We assign a weight to each hypothesis as in a standard $\chi^2$ analysis,
integrating the distribution from the obtained value up to infinite,

\[
W_{\hat{M}} \equiv F(\chi^2_{n_{eff}} > U(X,\hat{M}) ) =  
\]
\begin{equation}
 = \int_{ U(X,\hat{M}) }^{+\infty} 
g(\chi^2_{n_{eff}}, n_{eff}(X,\hat{M}) ) d\chi^2_{n_{eff}} 
\label{pw}
\end{equation}
i.e., the probability of finding a value of $\chieff$ bigger than 
or equal to $U(X,\hat{M})$. Here, we explicitly write where 
the dependence in the data ($X$) and in the parameters ($\hat{M}$) is.

\section{Checking the method}

In this section, we will
test the whole method in the problem of
one parameter estimation, but first, 
we will study the quality of our 
approximation to the distribution function
of a $\chi^2$ with correlations.
These two points will be done by means of Monte Carlo
simulations. In order to do that, we have
chosen the JB-IAC 33 GHz Interferometer  
\citep{melhuish99} as the reference 
experiment.

This experiment is a two element interferometer, which 
operates at $33$~GHz, at the Teide Observatory. 
It has two configurations, with angular resolutions 
$2^o$ ($\ell = 106 \pm 19$), and $1^o$ ($\ell = 208 \pm 18$), 
respectively. The window function in both configurations is 
very narrow, so the results are quoted in terms of total 
power inside the band (band power).
The experiment has given measurements on the power spectrum 
on both scales \citep{dicker99,harri00}, which 
are consistent with the Boomerang data \citep{bern00}.
We have the likelihood analysis implemented for this 
experiment, so the comparison with the new method will 
be straightforward. 
In our analyses, we have used the compact configuration, 
and only one of the two channels (i.e., the real part 
of the complex visibility).

The CMB realizations have been done assuming the 
following values for the cosmological parameters: 
$n =1$, $\Omega = 1$, $\Omega_b  = 0.03$, 
$\Omega_\Lambda = 0.7$ and 
$H_0 = 75 km\; s^{-1} Mpc^{-1}$. 
For this model, the total power inside the window function 
(or band power) is $BP = 51.45 \mu K$ for the short 
configuration ($\ell = 109 \pm 18$).
This number is related with $\sigma_{sky}$ 
by a conversion factor, which 
is obtained using the flat 
band power approximation 
(i.e. $BP = \ell (\ell +1) \cl /2\pi$ 
constant inside the window function)
in equation (\ref{def_ssky}).
For our instrument, this conversion factor is 
$BP = 5.44 \sigma_{sky}$, 
which gives $\sigma_{sky} = 9.46 \mu K$ for the previous model.
For this experiment, the sensitivity in a 
$30$s integration is given by 
$\sigma_{noise} \approx 250 \mu K / \sqrt{N_{days}}$, 
where $N_{days}$ is the number of observing days. 
Therefore, the signal-to-noise ratio is given
by $w = \sigma_{sky}^2/ \sigma_{noise}^2$.

\subsection{Our approximation to the distribution function}

\begin{figure}
\includegraphics[width=\columnwidth]{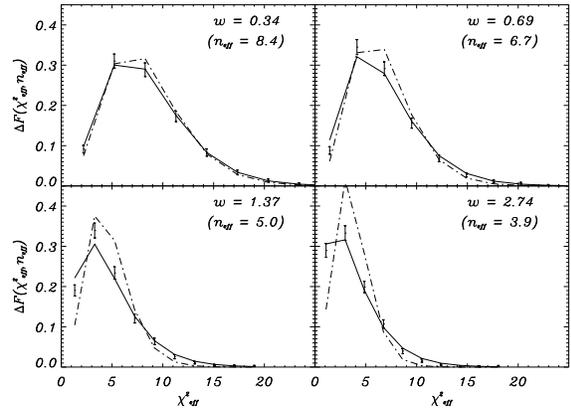}
\caption{Distribution function for a $\chi^2$ with correlations 
with $N=10$ terms. We show the histogram with the frequencies 
for the rescaled $\chi^2$, obtained from $1000$ realizations, for 
four different cases, varying $w$ (signal-to-noise ratio). 
We use $10$ bins of equal size
to sample the distribution function. In all the figures, the dots 
represent the numbers coming from the realizations, 
and the error bars show their sampling error.
The solid line is our approximation to the distribution function, 
using the value of $n_{eff}$ from the formula (shown within 
parentheses in the figure). It is also shown, using
dot-dashed lines, the distribution function of a (rescaled) 
$\chi^2$ with $N=10$. We can see the effect of the correlations 
as $w$ increases.}
\label{fig_histog}
\end{figure}

The first point is to check the validity of our 
approximation to the distribution function for 
a $\chi^2$ with correlations.
We will study the case when the $x_i$ quantities
entering in the $\chi^2$ follow a multivariate 
gaussian distribution.
This is the case for an (ideal) CMB map, where 
the temperature at each pixel has two contributions,
one coming from gaussian noise, and 
another one from the cosmological fluctuation 
field, which is supposed to be also gaussian.
Using the CMB terminology from Section 2, we will study the 
distribution of the statistic

\begin{equation}
\chi^2 = \sum_{i=1}^{N} \frac{ x^2_i }{ \sigma^2_i + \sigma^2_{sky} } 
\label{chi3}
\end{equation}

\noindent
and we will compare it with the proposed approximation.
This is a particular case of (\ref{chi2}), when 
$P_i = 1$. We will study in detail this case here,
but our results are completely general.
It should be noted that our proposal is exact, 
by definition, in the two limit cases of no correlations
at all, and totally correlated points.
The first one correspond to the definition of the $\chi^2$
distribution function, and the second one is the case 
of a $\chi^2$ with $N=1$.

In Appendix C we present the formal aspect of the 
distribution function for (\ref{chi3}), and we study in 
detail, analytically, the case $N = 2$.
The cases with low $N$ turn out 
to be the critical ones, because the
shape of the distribution function differs strongly 
from a gaussian.
In the limit of high $N$, both our approximation 
and the real distribution function tend to a 
gaussian distribution (the same one, by definition),
as a consequence of the Central Limit Theorem.
Therefore, it is interesting to test our proposal 
for a intermediate range of values of $N$.
We have done so, and we will present here, as an 
example, the case for $N=10$.

We generate CMB realizations with noise, 
for a ten pixels map.
From each simulation, we compute (\ref{chi3}), and from 
the whole set of values obtained, we study the histogram 
with the frequencies, and we compare it with the proposed 
one, for several values of the 
signal-to-noise ratio. We show these results in
Figure \ref{fig_histog}.
In general, the asymptotic shape of 
the distribution is very well
reproduced, and the largest differences always occur
for low values of $\chi^2$. This is precisely 
the kind of approximation we need, 
because in a statistical analysis we usually are 
interested in the tail of the distributions.

We can see that the distribution of the $\chi^2$ with 
correlations among terms is compatible, inside the 
numerical precision, with an standard 
(rescaled) $\chi^2$, with 
an effective number of degrees of freedom, 
smaller than $N$.

We have also checked that the numerical values for $n_{eff}$ and 
$A$ are correctly given by equations (\ref{neff}) and (\ref{A}).
In Table \ref{tabla2} we compare the values obtained from the
simulations with the predicted ones given by the theoretical formulae.
Their difference is in all cases smaller than 
the sampling errors, so we conclude that 
our expressions give the correct values 
for these parameters. 

\begin{table}
\caption{Values of $n_{eff}$ and $A$, for $N=10$.} 
\label{tabla2}
\begin{tabular}{@{}cccccc@{}}
\hline
\hline   
  & \multicolumn{2}{c}{Theoretical values$^{a}$}  & &  
\multicolumn{2}{c}{Numerical values$^{b}$} \\
\cline{2-3} \cline{5-6}
        $w$ &       $n_{eff}$ &       $A$ &  &  $n_{eff}$ &  $A$ \\ 
\hline
0.34 & 8.4 & 0.84 & & 8.4 & 0.84  \\
0.69 & 6.7 & 0.67 & & 6.8 & 0.67  \\
1.37 & 5.0 & 0.50 & & 4.8 & 0.47  \\
2.74 & 3.9 & 0.39 & & 3.8 & 0.37  \\
\hline
\hline
\end{tabular}

\medskip
$^a$ Computed using eqs. (\ref{A}) and (\ref{neff}).\\
$^b$ Computed using 100 CMB realizations, 
from the numerical value of $<\chi^2>$ and $MS(\chi^2)$.
\end{table}

\subsection{Applying the method to one parameter estimation}

We will now test our method in the problem of
one parameter estimation. The idea is to compare, by means 
of Monte Carlo simulations, our method with the Maximum 
Likelihood on the full map, which is widely accepted as 
the optimal method for CMB analyses. We will consider in 
detail the problem of determining the total power measured by  
a given experiment, so our parameter will be $\sigma_{sky}^2$.

Both the $\chi^2_M$ and the ML are, by construction, 
almost unbiased methods for 
determining the total power of an experiment. 
In order to compare them, we 
have studied the power of each one.
The power of an statistical method, when 
determining a certain parameter,  
is characterised by the $RMS$ of the estimate of this parameter.
The smaller this value, the more powerful 
is the method.
To compute it, we will use Monte Carlo 
simulations for a fixed CMB sky plus simulated noise. 
We will consider different values for the signal-to-noise ratio, 
and for each one, we will obtain the $RMS$ for each method.
Finally, we will also compute the 
degree of coincidence of both methods, 
which can be parameterised through 
the quantity $RMS(ML,\chi^2_M)$, 
defined as the RMS of the difference between their estimates.

For the experiment we are 
considering, we have generated simulations
for an observation of declination $+41^o$, and 
R.A. range $8h$-$18h$, for a fixed
CMB signal ($\sigma_{sky} = 9.46 \mu K$).
Each realization contains, for a single channel, 
$300$ data points,
with a pixel size of $0.5$ degrees.
Given that this experiment has a narrow window function, the 
band power is directly a measurement on the power spectrum.

The values of the $P_k$'s turn out to be not critical
in this problem. The solution to the equation
(\ref{P_k_ssky}) is $P_k \approx 1$, $\forall k$, so 
we use here $P_k = 1$.
The results of the realizations are summarised in 
Table \ref{tabla1}. We conclude that, in this problem,
the maximum likelihood on the full map and
the $\chi^2_M$ method have the same power, within
the uncertainty. 

\begin{table*}
\caption{Comparison of the power of the ML and 
the $\chi^2_M$ methods for a single parameter estimation: the total 
power measured by an experiment ($\sigma_{sky}^2$). The values were obtained from
Monte Carlo simulations of the model 
$\sigma_{sky} = 9.46 \mu K$, which corresponds to 
$BP = (51.45 \mu K)^2 = 2647 \mu K^2$.
We quote the results in terms of band power ($BP$).} 
\label{tabla1}
\begin{tabular}{@{}cccccccc@{}}
\hline
\hline
 &  &   & \multicolumn{2}{c}{$\chi^2_M$ method} &  & ML method & \\  
\cline{4-5} \cline{7-7} \\
$N_{days}$ & $\sigma_{noise}$  & $w$ & $n_{eff}$ & 
$RMS( E[\sigma_{sky}^2] )^a$ &  & 
$RMS( E[\sigma_{sky}^2] )^a$&  
$RMS(\chi^2_M,ML)^b$ \\        
  &       $(\mu K / pixel)$  &  &  &  $(\mu K^2)$ &  & $(\mu K^2)$  
&  $(\mu K^2)$ \\
\hline
  90 & 26.35 & 0.13 & 7.27 & 1170 &  & 1150 & 590 \\
 120 & 22.82 & 0.17 & 6.94 &  980 &  & 1065 & 520 \\
 240 & 16.14 & 0.34 & 4.50 &  800 &  &  805 & 410 \\
 480 & 11.41 & 0.69 & 3.43 &  640 &  &  680 & 410 \\
 960 & 8.07  & 1.37 & 2.95 &  570 &  &  640 & 420 \\
1920 & 5.71  & 2.74 & 2.67 &  530 &  &  570 & 440 \\
3840 & 4.03  & 5.51 & 2.65 &  500 &  &  540 & 490 \\
\hline
\hline
\end{tabular}

\medskip
$^a$ We report here the $RMS$ for both methods, obtained 
from $500$ CMB plus noise realizations, for different signal-to-noise
ratios ($w$). The sampling uncertainty in the 
RMS can be obtained as $\sigma_{RMS}^2 = 2 RMS^2 / N_{realizations} $. 
We can see that the values coming from both methods are compatible, 
inside that uncertainty. We have also checked that both methods give 
unbiased estimates, i.e. $< E[\sigma_{sky}^2] > = \sigma_{sky}^2$.
We do not include the values for $< E[\sigma_{sky}^2] >$ in the table
for clarity.\\
$^b$ We report here the degree of coincidence of both 
methods, i.e. the quadratic dispersion of the estimates from
both ones. The values were obtained from $12$ CMB plus 
noise realizations.\\
\end{table*}

If we study the degree of coincidence of both
methods, by computing the quantity $RMS(ML,\chi^2_M)$, we
conclude that both methods are highly coincident 
when the signal to noise is low. 
For example, 
for $w=0.13$, the $RMS$ of both methods is $\sim 1150 \mu K^2$, 
and the dispersion between estimates is $590 \mu K^2 $, 
roughly half the $RMS$. So, not only the power of 
the methods is similar for low $w$, but also the estimates are.
For high values of $w$ ($w \ga 0.6$), both methods tend to be 
independent, in the sense that the degree of coincidence $RMS(ML,\chi^2_M)$
approaches to the value of the $RMS$.

Once we have the (approximated) distribution function, 
we can determine the confidence region for one particular 
experiment. Usually, the size of this region is given by 
the $68 \%$ of the probability. Our error bars has to be 
interpreted in a frequentist way. 
That is, if we make lots of simulations, the
true signal will lie inside the confidence region of each 
realization the $68 \%$ of the times. 
It is known that this value does not 
necessarily represent the 
'error bars' defined in the usual Bayesian way, 
by treating the band-power likelihood function as a 
probability distribution.

As an example of application of our method, 
we have analysed the data 
from Dicker et al. (1999). These data correspond to 
declination $+41^o$, observed with the low resolution 
configuration ($\ell = 106 \pm 19$).
The value obtained using the likelihood analysis is 
$\Delta T = 43 ^{+13} _{-12} \mu K$.
In order to compute the $C'$, we use a value for 
$\Delta T = 40 \mu K$. In any case, our result does not 
depend on this initial value, and starting
with another one 
($\Delta T = 20 \mu K$ or $\Delta T = 60 \mu K$, for
example) gives the same result in the first iteration.
The estimated value using our method,
with $P_k = 1$, is 
$\Delta T = 43 ^{+9} _{-11} \mu K$, where the C.L. are 
defined as the $68 \%$. The effective
number of degrees of freedom is this case 
was $n_{eff} = 18.14$, 
and the total number of points where the CF was evaluated
was $n = 20$. The lag used for sampling 
the CF was $0.67$ deg, but the result is not sensitive to changes
in this number.

We have obtained the same estimate as the 
likelihood, but our confidence region is smaller. 
As we have pointed before, this probably is due 
to the fact that the confidence region 
has a different definition in both methods. 
In order to compare those confidence levels, 
we perform Monte Carlo simulations, 
using the measured signal and the experimental noise, and 
we obtain the equivalence between  
the confidences levels for both methods. We conclude 
that the region that contains the $68 \%$ of the area of the 
likelihood around the peak, corresponds to $\sim 75 \%$ of the 
probability in a frequentist sense (i.e., that region contains 
the true signal the $\sim 75 \%$ of the times).
This explains why the likelihood give us bigger error
bars in this particular problem.
In a general case, we will have to repeat this analysis for the 
likelihood estimate, in order to compare the sizes of
the confidence levels.

\section{Estimating several parameters}

In the previous section, we have proved 
that our method, when constructed from
the CF, has the same power as the likelihood when determining 
a single parameter (an overall normalisation).
We now probe if this is true for a larger number of parameters. 

For the case of several parameters estimation, 
the CF has proved to be a very good statistic 
in determining the power spectrum of the CMB \citep{szapudi01}.
In their paper, they obtain the $\cl$'s 
by a Gauss-Legendre integration of  
the CF. 
When applied to simulations of the Boomerang data \citep{bern00}, 
the error bars for that method, coming from 
Monte Carlo simulations, are of the same order 
as the sample variance, which is the theoretical limit to 
the size of the error bars.

Here, we will apply our method to the COBE
Differential Microwave Radiometer data in order to 
obtain the $\cl$'s in two cases. The first one,  using the 
power law parameterisation of the angular power spectrum, 
in terms of the spectral index of the primordial spectrum, $n$ 
(with $P(k) \propto k^n$), and the normalisation parameter, $Q_{rms-PS}$. 
In this case, using our notation, we have $\hat{M} = \{ n, Q_{rms-PS}  \}$.
The dependence of the $\cl$'s in these parameters, for
a pure Sachs \& Wolfe spectrum (which dominates in the considered
multipole range), is given by \citep{bond87}

\begin{equation}
\cl = \frac{4\pi}{5} Q_{rms-PS} 
\frac{ \Gamma( \ell + \frac{n -1}{2} ) \Gamma(\frac{9-n}{2}) }
{ \Gamma( \ell + \frac{5-n}{2} ) \Gamma( \frac{n+3}{2} ) }
\end{equation}

\noindent
The second case will be to estimate several $\cl$'s directly, i.e., 
$\hat{M} = \{ \cl \} _{\ell=\ell_1} ^{\ell=\ell_p}$.

\subsection{Estimating ($n, Q_{rms-PS}$) for the COBE data}

The CF has been applied for COBE analyses of the $(n,Q_{rms-PS})$ 
parameters by other authors. In \cite{hinshaw96b}, the 
quadrupole normalisation is inferred using 
Monte Carlo-based gaussian likelihood analysis for a 
scale-invariant ($n=1$) power-law spectrum. 
In \cite{bunn94} (hereafter BHS94), they determine 
$n$ and $Q_{rms-PS}$ as the best fit values to the computed CF.
They show, by means
of Monte Carlo simulations, that this method is not optimal
for the determination of those parameters, because
they obtain a large $RMS$ when trying to recover them
simultaneously (see Table 2 in that paper). 
Nevertheless, other estimators, such as the direct evaluation 
of the $a_{\ell m}$, give smaller $RMS$'s.
We will use here the CF, but with our method, to probe whether 
we obtain good results.

In order to test our method with these two parameters, we perform 
Monte Carlo realizations of COBE like maps, using a scale invariant power 
spectrum ($n=1$) with a normalisation 
$Q_{rms-PS} = (18 \mu K)^2 = 324 \mu K^2$.
We will use the standard COBE pixelization of 
$N=6144$ pixels (index level 6), in galactic coordinates.
We include in our realizations the noise level corresponding to 
the combined 4-year COBE map of the three frequencies, using the weights 
quoted in \cite{hinshaw96b} (this map is referenced there as 
$31+53+90$).
The CF has been sampled using a $2.6^o$ step, 
as it appears in that paper (that size correspond to the typical
pixel size).
Anyway, we have checked that our results are consistent 
when changing that step.
In our calculation of the CF,
we use a galactic cut $|b| > 20^o$. 

\begin{table*}
\caption{Results of Monte Carlo simulations for the determination
of $n$ and $Q_{rms-PS}$ with the $\chi^2_M$ method, using 
different sets of $P_k$ parameters. We explore the
cases of: (a) an uniform value of $P_k$ for $k=1,...,kmax$ and zero
the others (first four rows, quoted as 'Uniform, kmax'); and 
(b) the optimum set of $P_k$ values. 
The MC simulations have parameters $n=1$ and $Q_{rms-PS}^{1/2} = 18
\mu K$, and the correlation function is sampled at $70$ equally spaced 
bins of size $2.6^o$. We use the galactic 
cut $|b| > 20^o$, and the noise levels of the 
combined 4-year COBE map.}
\label{tabla3}
\begin{tabular}{@{}cccccc@{}}
\hline
\hline   
$P_k$ $^a$ & $n$ $^b$  & $Q_{rms-PS}$ $[\mu K^2]$ $^b$ & 
$n_{eff}$ & $RMS(n)$ $^c$ & $RMS(Q_{rms-PS})$ $[\mu K^2]$ $^c$ \\
\hline
Uniform, 70 &  $1.04 \pm 0.54$ & $323 \pm 184$ & $2.55$ & 0.60 & 190 \\
Uniform, 41 &  $1.02 \pm 0.50$ & $331 \pm 178$ & $1.92$ & 0.52 & 177 \\
Uniform, 10 &  $1.08 \pm 0.34$ & $307 \pm 147$ & $1.12$ & 0.28 & 133 \\
Uniform,  3 &  $0.99 \pm 0.45$ & $327 \pm 145$ & $1.05$ & 0.34 & 147 \\
\hline
Optimal  &  $1.03 \pm 0.28$ & $316 \pm 141$ & $1.78$ &  0.19 & 114 \\
\hline
\hline
\end{tabular}

\medskip
$^a$ Adopted values for the $P_k$'s. 
The last row is the optimum set of $P_k$ values 
(see Figure \ref{figura2}). \\
$^b$ Results from 100 Monte Carlo simulations of COBE
data (see details in the text). 
The first number is the average value for the parameter, and 
the second one is the $RMS$ from the simulations. \\
$^c$ $RMS$ values obtained analytically, using the linear approximation 
to the CF (see Section 3).
\end{table*}

We will use the $P_k$ quantities given by 
the minimum of the function
$RMS(E[n]) \times RMS(E[Q_{rms-PS}])$, as we have discussed 
in Section 3.
Those $RMS$'s can be derived, using the linear approximation
to the CF, from equation (\ref{rms_several}).
In this problem, the $R$ and $J$ quantities are
given by

\begin{equation}
R = \Bigg( \sum_i \frac{P_i f_{i,1}^2}{\sigma^2(C(\theta_i))} \Bigg)
\Bigg( \sum_i \frac{P_i f_{i,2}^2}{\sigma^2(C(\theta_i))} \Bigg) - 
\Bigg( \sum_i \frac{P_i f_{i,1}f_{i,2} }{\sigma^2(C(\theta_i))} \Bigg)^2
\end{equation}

\begin{equation}
J_{k,1} = f_{k,1} \Bigg( \sum_{j} \frac{P_j f_{j,2}^2}{\sigma^2(C(\theta_j))} 
\Bigg) 
- f_{k,2}\Bigg( \sum_j 
\frac{P_j f_{j,1}f_{j,2} }{\sigma^2(C(\theta_j))} \Bigg)
\end{equation}

\noindent
where $1$ stands for $n$, and $2$ for $Q_{rms-PS}$.
The equation for $J_{k,2}$ can be obtained 
from $J_{k,1}$, just interchanging $1 \leftrightarrow 2$.

In order to check the previous expressions for the $RMS$, 
we use 100 of the above mentioned realizations of 
COBE like maps ($n=1$; $Q_{rms-PS}^{1/2} = 18 \mu K$), 
and we analyse them using several sets of $P_k$'s. 
In this way, we can obtain
the real value of the $RMS$, and compare it with the number 
coming from the formula.
The results are summarised in Table \ref{tabla3}.
In all cases, the theoretical numbers obtained from
equation (\ref{rms_several}) are in agreement 
with the numerical results, so
we conclude that the linear approximation 
to the true CF works well in computing 
the $RMS$. 
The largest differences occur when we obtain a large $RMS$, due 
to the fact that, in that case, fails the linear approximation
to the CF.

The average values recovered for $n$ and $Q$ from Monte-Carlo simulations 
show that the estimator is unbiased, as we would expect. 
It should be noticed that, 
in Table \ref{tabla3}, the effective 
number of degrees of freedom is quite small.
In all cases, we obtain $n_{eff} \la 3$, but 
we are using $n = 70$.
The reason is that the CF contains long-range
terms, coming from low multipoles ($\ell \sim 2$).
This fact reduces the degrees of freedom drastically, 
so the choice of the $P_k$ will be critical in this 
problem. 
The numbers obtained when we consider the whole CF
and $P_k = 1$ are 
compatible with those in BHS94, but slightly
better because we consider the noise of the
4-year COBE map. In that paper, they obtained, 
using the same galactic cut ($|b| > 20^o$), and 
the noise from the 2-year map, the values
$RMS( E[n] ) = 0.96$ and 
$RMS( E[Q_{rms-PS}]) = 253\mu K^2$ (in our units).
Nevertheless, we see that considering only the
first points, and setting to zero the others, 
strongly reduces the $RMS$ of the estimate, 
even below the values obtained 
when they do not consider noise and incomplete 
sky coverage (they have 
$RMS( E[n] ) = 0.36$ and 
$RMS( E[Q_{rms-PS}] ) = 175 \mu K^2$).

\begin{figure}
\includegraphics[width=\columnwidth]{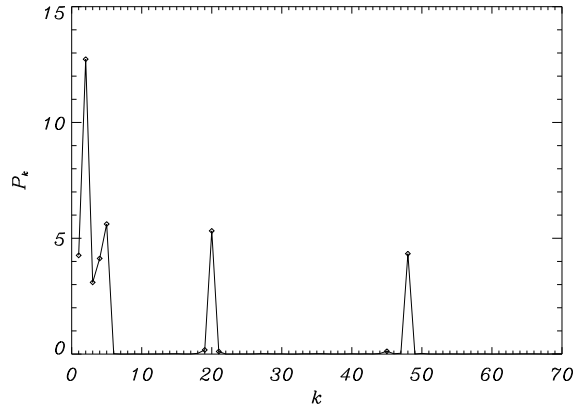}
\caption{Optimum set of $P_k$'s for the $31+53+90$ 4-year {\sl COBE} map
(see details in the text), using the galactic 
cut $|b| > 20^o$. These values were obtained by numerical
minimisation of the product 
$RMS( E[n] ) \times RMS( E[Q_{rms-PS}] )$, 
using the linear approximation to the CF. Each value for $k$ 
($1 \le k \le 70$) corresponds to an angular distance, 
given by $\theta_k = (k-1) 2.6^o$. An explanation of this peculiar 
structure can be found in the text (Section 6.1).}
\label{figura2}
\end{figure}

Finally, we can obtain the optimum set of 
$P_k$'s by numerical minimisation of the product
of $RMS$'s. We have used the Fortran program
$amoeba$ \citep[p.402]{numerical}.
The obtained value for these quantities 
is shown in Figure \ref{figura2}.
These numbers do not depend on the initial guess for 
$n$ and $Q_{rms-PS}$ within the a priori region of uncertainty.
The values obtained have a peculiar form, but
it can be understood as follows.
We see that the terms which contribute to 
the optimum estimator (minimum $RMS$) 
correspond to the first $\sim 10$ degrees, i.e.
the first part of the CF, as we would expect.
But there are also two regions,  
one at $\sim 50\degr$,  and 
another at $\sim 125^o$, which contribute to the estimator.
These peaks are located just in the zeros of the
quadrupole ($\ell =2$) which is the multipole with 
the largest cosmic variance. By using those points, we add
some information to the estimator (coming from higher 
$\ell$ values), but we do not increase its variance.
In any case, to consider or not those points do not affect
too much to the power of the method (in Table \ref{tabla3}, the values
obtained when using the first 10 points from the CF are 
close to those obtained with the optimal set). 
Following this interpretation, one could think about other combinations
of the $P_k$ parameters using others points (for example, taking
two points at $\sim 40\degr$ and $\sim 105\degr$. We have explored this
possibility as an illustration, and we find the values $RMS(n)=0.31$
and $RMS(Q_{rms-PS})=144~\mu K^2$, which are similar but slightly higher
than those obtained for our optimal $P_k$. 
Thus, we can conclude that for this problem, we can find a set of estimators
(one for each one of these sets of $P_k$ values) 
which give similar RMS values when estimating $n$ and $Q$, and
as we show below, these values are comparable to those given
by the likelihood method. 

Finally, we have applied our $\chi^2_M$-test 
for the CF to the actual 4-year COBE data. 
Our estimates, from the analysis of the
$31+53+90$ map, using the galactic cut $|b| > 20^o$, 
a step of $2.6^o$ to sample the CF, and
the optimum set of $P_k$'s, are 
$E[n] = 1.08$, and $E[Q_{rms-PS}^{1/2}] = 15.2 \mu K$, 
so our estimate of the parameters, using the 
true $RMS$ from $100$ realizations, will be 
$n = 1.08 \pm 0.28$, and 
$Q_{rms-PS}^{1/2} = 15.2 \pm 3.5\mu K$.
This result has to be compared with the likelihood analysis 
using these data (see \cite{hinshaw96a}, Table 1).
They obtain for this map 
$n = 1.25^{+0.26}_{-0.29}$, 
$Q_{rms-PS}^{1/2} = 15.4 ^{+3.9}_{-2.9}\mu K$.
The estimates from both methods 
are nearly the same, and now 
the error bars coming from the CF are compatible in size with
those coming from the maximum likelihood method.
For comparison, using no weights ($P_k = 1$), we would obtain 
$E[n] = 1.10 \pm 0.54$, and 
$E[Q_{rms-PS}^{1/2}] = (16.0 \pm 4.8) \mu K$, so we can see
that using the $P_k$'s for this problem is essential.

\subsection{Estimating band power spectra for COBE data}

Finally, we will apply our method to obtain band power
spectra for the COBE data. We have used the realizations from the 
previous subsection ($n=1$; $Q_{rms-PS}^{1/2} = 18 \mu K$), 
and the same COBE map. We will compare our results 
with those from \cite{hinshaw96a} (see Table 2 in that paper).
We have used exactly the same $\ell$ range: four multipole bands 
between $\ell =2$ and $\ell = 40$.
Those bands are: $2 \le \ell \le 5$; $6 \le \ell \le 10$; 
$11 \le \ell \le 20$ and $21 \le \ell \le 40$.
Using those realizations, we have checked that the method is unbiased
when applied to power spectrum estimation.

When applied to the $31+53+90$ 4-year COBE map, we obtain  
the results that are shown in Table \ref{tabla4}.
We quote the band power values in terms of the 
quadrupole normalisation expected for a scale-invariant 
power-law spectrum within the specified range of $\ell$.
The quoted values for our method have been obtained from 
the optimal set of $P_k$ for this problem.
When compared with the ML data, we can see that the error
bars are of the same order in both cases, as in the previous subsection.
So we again obtain a method of a similar power to the likelihood on 
the pixel map.

\begin{table}
\caption{Band power values for the COBE data 
(31+53+90 map)$^a$.} 
\label{tabla4}
\begin{tabular}{@{}ccccc@{}}
\hline
\hline   
 & \multicolumn{4}{c}{Multipole Band} \\
\cline{2-5} \\ 
Method & $2 \le \ell \le 5$ & $6 \le \ell \le 10$ &
  $11 \le \ell \le 20$ & $21 \le \ell \le 40$ \\
\hline
$\chi^2_M$, optimal  & $17.0 \pm 3.3 $  & $9.0 \pm 2.6$  
	& $17.6 \pm 1.8$  &  $0 \pm 4.4$  \\
ML $^b$ & $18.6^{+4.5}_{-3.4}$ & $16.7^{+2.4}_{-2.0}$ 
	& $20.3^{+2.2}_{-2.1}$ & $1.0^{+13.2}_{-1.0}$ \\
\hline
\hline
\end{tabular}

\medskip
$^a$ These band power amplitudes are expressed in terms
of $Q^{1/2}$, i.e. the quadrupole normalisation expected for a scale-invariant
power-law spectrum within the specified range of $\ell$. The units are
$\mu K$. \\
$^b$ ML values from \cite{hinshaw96a} for the same map.
\end{table}

The estimates from both methods are consistent in all bins, except
the apparent inconsistency at the second one, where the two
estimates differ in more than 3 times the size of one error bar.
Nevertheless, the statistical significance of that deviation has to be
computed as follows: given that we do not know the true value
for the band power, when comparing two results we have to consider the
difference of both estimates, and so we have
to compute the variance of the difference. In this case, the difference is 
$16.7\mu K - 9\mu K = 7.7 \mu K$, and the $RMS$ of the difference of those
two estimates is $RMS \sim \sqrt{ 2.6^2 + 2.2^2} \sim 3.4\mu K$.
Here we are assuming the fact that both methods are almost independent, 
following the results for the case of 
one parameter estimation, with signal-to-noise ratios of the order of $1$.
Therefore, we find a deviation at the $2.3$-sigma level, which 
corresponds to a fluctuation of 1 in 47 (for a normal distribution).
This can be understood given that the ML and the $\chi^2_M$ are two different
methods: the first is based on the full map, and the second on the
correlation function. Therefore, the estimates will be different in general,
although as we can see, both methods have similar power, so there 
is no reason to consider any one of them as ``the estimate''.

Summarising, we have seen that it is possible with our method to perform an 
analysis with a similar power to the ML, 
even for the case of several parameters.

\section{Discussion and conclusions}

In this paper we have presented an statistical method to analyse CMB
maps. It consists in a variation of an standard $\chi^2$-test for 
the case when we have correlated points. 
Here, our test has been constructed from the two-point correlation
function, following the proposal from other authors \citep{bashinsky01}
that the CF contains all the relevant information concerning
the cosmological parameter estimation. In this line, 
we propose a $\chi^2_M$ based on the CF, which
is a different approach from the ``usual $\chi^2$''
method (which uses the full covariance matrix $C'$). 
Our proposal explicitly uses only the
diagonal terms of the covariance matrix of the data, but we introduce
certain weights, which now implicitly contain all the correlations.
This approach has two important 'computational' advantages compared 
with the ``usual $\chi^2$'', or with a likelihood based on the CF: 

\begin{itemize}
\item we do not need to invert the covariance matrix, $C'$; all the 
quantities (the $\chi^2_M$, $n_{eff}$, $A$ and the expression for 
the $RMS$) depend on $C'$ directly.
\item even more, if we have a problem with a low value for $n_{eff}$, 
the effective number of $P_k$ to use (i.e. the number of $P_k$'s which are
significantly different from zero) will be also small, typically, of
order $O(n_{eff})$. So it is not even necessary to use
all the terms of the diagonal.
\end{itemize}

These advantages are more important 
when comparing our method with the standard
ML analysis on the full map. 
We do not need to invert the covariance
matrix of the map ($N \times N$), and we only need to concentrate on
a few numbers, so the problem is computationally accessible if
we have to deal with large datasets.  
It it important to stress here that 
the $\chi^2_M$ is not an approximation to
the ML on the full map, but a different approach 
(i.e. both methods will give different estimates and 
probability contours for a given problem). So the $\chi^2_M$  
can be applied to any problem, but it has to be checked, by means of 
Monte Carlo simulations, that the method has a similar power to 
the maximum likelihood. As we have seen, this is the case for several CMB
common problems (power spectrum estimation, and cosmological parameter
estimation).

The largest computational effort in our method has to be done estimating 
the CF, which is a $\sim N^2$ operation. Nevertheless, there are estimators
for the CF more efficient \citep{atrio01,szapudi01b}, 
so our procedure could be applied
to current {\sl WMAP} data, and {\sl PLANCK} simulated
data, but this will be treated in detail in future works.

Our method can be extended for general noise covariance 
matrices. We only need to compute $C_N (\theta)$ in the
same way as the CF, and introduce it in equation (\ref{chi2}).
If we have the noise matrix, it is straightforward to 
obtain $C_N (\theta)$. But, if we do not have the noise matrix, 
we can obtain an estimate of the correlation function
of the noise using MC simulations.
This idea has been used recently by other authors 
\citep{szapudi01b}. It must be noted that if the noise
changes substantially from pixel to pixel, then 
we would have to use weights in equation (\ref{est_cf}) to 
compute the CF in a more efficient way.

We have also presented an approximation which gives very accurately 
the distribution function for a $\chi^2$ constructed from 
a set of multivariate gaussian variables.
This proposal can be extended to approximate the distribution
function of any quantity made of a sum of squares, each of them 
distributed (exactly or approximately) following a gaussian distribution.

To conclude, we propose that, if we are interested in obtaining a certain 
set of parameters, $\hat{M}$, we can use an unbiased estimator of
certain quantities depending on these parameters, provided that they
contain all the relevant information to these parameters (in our case,
we have used the CF at certain angles). 
From it, we may obtain,
by varying the $P_k$'s, the best estimator of those
parameters.
In this paper, we have tested this proposal, using the
two-point CF as the reference estimator, in CMB problems.
For the case of one parameter estimation (the normalisation 
of the spectrum), our method, with the CF,  
turns out to be as powerful as the ML. 
When applied to COBE data, we have shown the importance of choosing
the right set of $P_k$'s. 
In the optimum case, we obtain a value for 
the $RMS$ two or three times smaller than 
the one obtained without
weighting at all. 
In this problem, the $P_k$'s are critical
because the effective number the degrees of freedom
is very small. The reason is that 
when we have strong correlations ($n_{eff}$ small compared with $n$), 
the structure of these correlations, 
which is encoded in the $P_k$'s,  will be relevant, 
so there will be a considerable difference between 
the optimal weights and $P_k=1$.
When analysing CMB data 
which contain large scales (low multipoles), 
we are considering correlations 
over long distances. All the points 
are correlated with the others due to these low multipoles 
(quadrupole, octupole,...). 
In the case of Boomerang data \citep{szapudi01}, the 
effective number of degrees of freedom will be larger
(if we throw away the large scales), so the $P_k$'s will be
closer to 1, and a standard $\chi^2$ test based on the CF
should produce good results.

\section*{Acknowledgments}

We would like to thank M.P.Hobson, R.A.Watson, R.Rebolo,  
C.Guti\'errez and R.Kneissl for their useful comments, and 
B.Barreiro for her COBE-like maps simulation program.

\appendix

\section{Relationship between the usual $\chi^2$ and the $\chi^2_M$ for 
gaussian linear problems}

In this Appendix we will study the relationship between the usual 
$\chi^2$ analysis (the standard approach for the case of correlated
data points, which uses the full covariance matrix), 
and our approach, for gaussian problems in which the model 
depends linearly on the parameters, and with a covariance matrix 
independent of the parameters. 
This case is not exactly equal to the usual one in CMB 
analyses, but it is very 
close and is particularly suitable for illustration. 

We will use here the following notation. Let $\vy$ = $(y_1,...,y_n)$ be 
a $1 \times n$ matrix containing the $n$ data points, which, by hypothesis,  
are distributed following a multivariate gaussian distribution.
Let $\valp$ = $(\alpha_1, ..., \alpha_k)$ be a $1 \times k$ matrix
whose elements are the $k$ parameters of the model. 
Let $\mx$ be a $k \times n$ matrix, also given by the model. Their elements
are defined so that the mean value of $\vy$, $< \vy >$, is given by the 
matrix multiplication $\valp \mx$.
Finally, let $\mM$ be the covariance matrix of the problem, defined 
as 
\[
\mM = < ( \vy - \valp \mx )^T ( \vy - \valp \mx ) >
\]
where $T$ stands for the transpose. 
Using the previous definitions, the usual $\chi^2$ and 
our $\chi^2_M$ are given, respectively, by

\begin{equation}
\chi^2   = ( \vy - \valp \mx ) \mM^{-1} ( \vy - \valp \mx )^T
\end{equation}
\begin{equation}
\chi^2_M = ( \vy - \valp \mx ) \mV^{-1} ( \vy - \valp \mx )^T
\end{equation}

\noindent
where we have defined the matrix $\mV$ using the diagonal of the 
covariance matrix, and our weights ($P_i$), in the following way: 
$\mV$$_{ii} = M_{ii} / P_i$, for $i=1,..,n$, 
and $\mV$$_{ij} = 0$ for $i \ne j$. It should be noted that 
the $\mV$ matrix depends implicitly on the weights ($P_i$).

For this family of models under consideration, the
optimum estimator is the maximum likelihood, which is given by

\begin{equation}
\like \propto \frac{ \exp (- \frac{1}{2} \chi^2)}{ det(\mM)} 
\end{equation}

Until this point, we have not made use of the fact that 
the covariance matrix is independent of the parameters. If 
we use it now, from the last equation we have that 
the maximum likelihood reduces to the usual $\chi^2$ for this problem. 
The estimate for both methods is obtained by minimising the previous 
expressions with respect to the parameters, so we have

\begin{equation}
E_1[ \valp ] = ( \vy  \mM^{-1} \mx^T )  ( \mx  \mM^{-1} \mx^T )^{-1}
\label{estimate1}
\end{equation}
\begin{equation}
E_2[ \valp ] = ( \vy  \mV^{-1} \mx^T )  ( \mx  \mV^{-1} \mx^T )^{-1}
\label{estimate2}
\end{equation}

\noindent
Hereafter in this section, we will use subscript $1$ for the standard
(usual $\chi^2$) method, and $2$ for the $\chi^2_M$. 
We compute now the covariance matrices of
the parameters, $\mW$, which are given by

\begin{equation}
\mW _1 \equiv Variance( E_1[ \valp ] - < E_1[ \valp ] > ) = 
( \mx  \mM^{-1} \mx^T )^{-1}
\end{equation}
\[
\mW _2 \equiv Variance( E_2[ \valp ] - <E_2[ \valp ]> ) =  
\]
\begin{equation}
 = ( \mx  \mV^{-1} \mx^T )^{-1} ( \mx \mV^{-1} \mM \mV^{-1} \mx^T ) 
( \mx  \mV^{-1} \mx^T )^{-1}
\end{equation}

\noindent
Using this notation, the $MS$ for each parameter is given
by the corresponding element in the diagonal of $\mW$.

\begin{figure}
\includegraphics[width=\columnwidth]{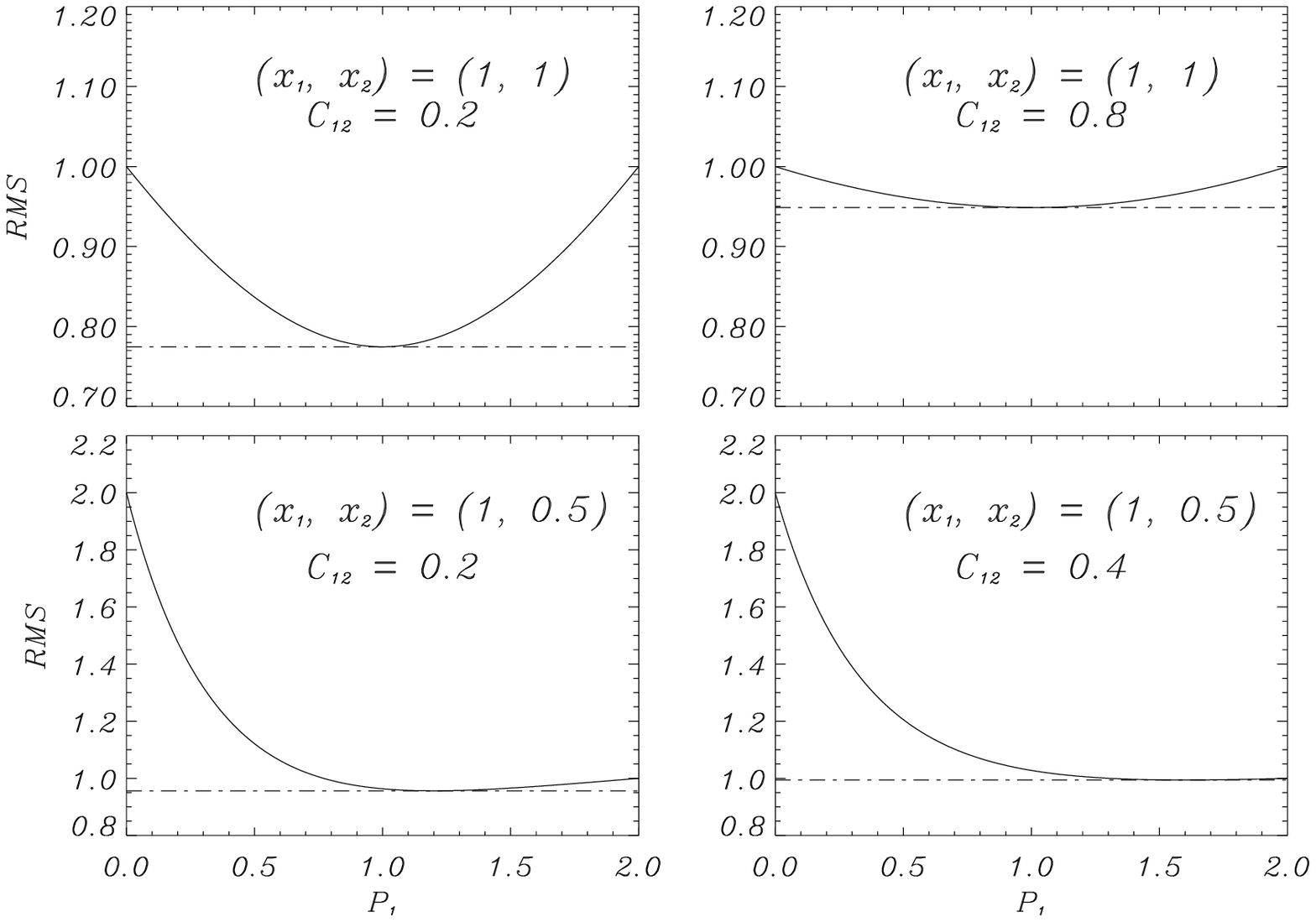}
\caption{RMS for the standard $\chi^2$ method with correlations,
and our $\chi^2_M$ method, for several linear gaussian models 
with only one parameter, $\alpha$, and $n=2$ variables, 
named ($y_1$,$y_2$), normally distributed with means $<y_1> = \alpha x_1$ and
$<y_2> = \alpha x_2$. 
We also assume a covariance matrix independent on the parameter, and
we parameterise it as  $\mM _{11}$ $= \sigma_1^2 = 1$, 
$\mM _{2}$ $= \sigma_2^2 = 1$, and $\mM _{12}$ $= \mM _{21}$ $= C_{12}$, 
but these results are completely general. We plot here four typical
cases for this problem. The dot-dashed line in all four
panels corresponds to the usual $\chi^2$ value for the RMS in the
estimate of $\alpha$. 
The solid line is the RMS obtained for the $\chi^2_M$, using the 
shown value of $P_1$. The $P_2$ is obtained from the normalisation 
equation  $P_1 + P_2 = 2$. 
We see that it is always possible to find a certain value of $P_1$, 
for which we have exactly the 
same RMS for both methods. For that $P_1$ value, the estimates
are also obtained to be equal (see text for details).}
\label{comparachi2s}
\end{figure}

The following point is to compare the estimates of both methods. 
For the case of one parameter estimation, it can be argued that
both method give the same estimate, and therefore have the same $RMS$.
The argument is as follows: in this case, the estimate for both 
methods is a linear combination of the $n$ quantities 
$\vy$. So it could be possible, in principle, to fix the 
$n$ quantities  ($P_i$) to equalise the $n$ 
coefficients in expressions (\ref{estimate1}) and (\ref{estimate2}).
Given that for this particular problem the usual $\chi^2$ is the
optimal method (i.e. the one with the minimum variance), and that
the corresponding estimator is the only linear one for which this variance 
is obtained, those 
$P_i$ quantities which set equal the coefficients in 
(\ref{estimate1}) and (\ref{estimate2}), are exactly the same that
would be obtained by minimisation of the $RMS$ of that parameter.
We have checked this statement for the critical case where we 
have a $\chi^2$ with only $n=2$ terms. 
In Figure \ref{comparachi2s} we present several particular examples, showing
that for the set of $P_i$ quantities that minimise the $RMS$ for the 
$\chi^2_M$, we always obtain the same $RMS$ as in the case 
of the usual $\chi^2$. We have also checked that, for those values 
of the $P_i$, the estimates from both methods are exactly the same.

Let's consider now the case of several ($k>1$) parameters. 
For this problem, it is not possible in general to obtain a set 
of $P_i$ quantities which render the $RMS$ of the usual $\chi^2$.
In any case, we have checked that it is always possible to choose 
those quantities to make the estimate for just one of the parameters 
exactly equal, and so its $RMS$.
We have studied this problem in detail for the case of $n=3$ terms
in the $\chi^2$, and $k=2$ parameters. The results are the following:

\begin{itemize}
\item if we choose to minimise the $RMS$ of only one parameter to find
the optimal $P_i$ quantities, we can always make the estimate for
that parameter exactly equal to the usual $\chi^2$. For those $P_i$ values,
the estimator for the other parameter is very close to the optimum, in 
the sense that the $RMS$ for the other parameter at the most $10\%$ bigger
that the optimal one.
\item if we minimise the product of the two $RMS$'s, we find in all cases 
that both estimates are close to the optimal ones, and the largest 
relative differences between $\mW_1$ and $\mW_2$ are smaller than $\sim 1\%$.
\end{itemize}

Therefore, we can conclude that the criteria to obtain the $P_i$ quantities
has an small ambiguity, in the sense that if we are interested
in one parameter in particular, we should minimise the $RMS$ for that 
parameter only. In practice, this ambiguity is not relevant because, for this
problem, the differences between the estimates and the obtained $RMS$ values
are negligible. Therefore, we will maintain the original proposal of
minimising the product of $RMS$'s, which in some sense is equivalent to
minimise the average size of the confidence region.
All these arguments can be applied to cases with $k>2$ parameters.

\section{ Useful quantities for multivariate gaussian distributions up to 
4th order}

The probability density function for a multivariate gaussian distribution
of n variables ($n-$MGD),  $X = \{ X_1, ...,X_n\}$, is given by 

\begin{equation}
g(X) = \frac{1}{ (2\pi)^{n/2} |C|^{1/2}} 
\exp{ \bigg( - \frac{1}{2} 
(X - \overline{X})^T C^{-1} (X - \overline{X}) \Bigg) }
\label{mgd}
\end{equation}

\noindent
where $C$ is called the {\sl covariance matrix}, $|C|$ stands for 
its determinant, and $\overline{X}$ is the mean of $X$. 
The elements of $C$ are given by 

\begin{equation}
C_{ij} = < (X_i - \overline{X_i})(X_j - \overline{X_j}) >
\label{cov_matrix}
\end{equation}

\noindent
We define $\sigma_i^2 \equiv C_{ii}$. When computing the mean square 
of equation (\ref{chi2}), or the $C'$ matrix given in (\ref{est_cf}), 
we need to know the following quantities: $< X_i^2 X_j^2>$, 
$< X_i^2 X_j X_k >$ and $< X_i X_j X_k X_l >$. We will obtain them here.
For a $2-$MGD, we can obtain the quantity

\[
< X_1^2 X_2^2>  = \int_{-\infty}^{+\infty} \int_{-\infty}^{+\infty} 
\frac{1}{2\pi \sqrt{|C|} } X_1^2 X_2^2 \times 
\]
\[
\times \exp{ \Bigg[ -\frac{1}{2|C|} (X_1^2 \sigma_2^2 + X_2^2 \sigma_1^2 
- 2 C_{12}X_1 X_2 ) \Bigg] } dX_1 dX_2 = 
\]
\begin{equation}
= \sigma_1^2 \sigma_2^2 + 2 C_{12}^2 
\end{equation}

\noindent
For a $3-$MGD, we obtain

\[
< X_1^2 X_2 X_3 > =  3 |C|^{2} 
( A_{22} A_{33} - A_{23}^2 ) 
( A_{11} A_{23} - A_{12} A_{13}) 
\]
\begin{equation}
 -2 |C| A_{23}
\end{equation}

\noindent
where $A_{ij}$ stands for the elements of the inverse 
of the C matrix. Using the same notation, 
for a $4-$MGD, we obtain

\[
< X_1 X_2 X_3 X_4 > = |C| \Bigg(2 A_{12} A_{34} - A_{14} A_{23} -  
A_{13} A_{24} \Bigg) - 
\]
\[
 - 3 |C|^{2} \Bigg( A_{12} A_{34}^2 + A_{14} A_{24} A_{33} -
- A_{14} A_{23} A_{34} - A_{12} A_{33} A_{44} - 
\]
\[
 - A_{13} A_{24} A_{34} + A_{13} A_{23} A_{44} \Bigg)
\Bigg( A_{11} A_{23} A_{24} + 
A_{12}^2 A_{34} -
\]
\begin{eqnarray}
- A_{11} A_{22} A_{34} + A_{13} A_{14} A_{22}  - A_{12} A_{14} A_{23} 
- A_{12} A_{13} A_{24}  \Bigg) 
\end{eqnarray}

\section{Probability distribution function for a $\chi^2$ constructed
  from  multivariate gaussian variables}

The moment generating function of the quadratic form 
$\chi^2 = \sum_{i=1}^{n} P_i X_i^2 / \sigma_i^2$, 
where the $X = \{ X_1, ...,X_n\}$ variables follow a $n-$MGD 
with zero mean (eq. \ref{mgd}), is given by (see \cite{mathai,kendall}) 

\begin{equation}
G(t) = \prod_{j=1}^{n} (1 - 2t\lambda_j)^{-1/2}
\end{equation}

\noindent
In this equation, 
$\lambda_1,...,\lambda_n$ are the eigenvalues of the 
matrix $\Sigma C$, where $C$ is the covariance 
matrix (\ref{cov_matrix}), and $\Sigma$ is defined as 
$\Sigma_{ij} = (P_i/ \sigma_i^2) \delta_{ij}$, being 
$\delta_{ij}$ the Kronecker-delta.
From this expression, the distribution function can be obtained by 
the Laplace inverse transform. Formally, we have 

\begin{equation}
\Psi_n(\chi^2) = L^{-1}[ G(-t) ]
\end{equation}

\noindent
where $L^{-1}$ stands for the inverse Laplace 
transform \cite[p.1142]{grad}, and we use 
the notation $\Psi_n(\chi^2)$ 
for the exact distribution function of the $\chi^2$.
As an example, 
we will study in detail the case $n=2$, comparing the 
exact distribution function with our approximation.

\subsection{Distribution function for n=2}

The analytical expression for the distribution function can be
obtained using \cite{grad}, eq. (17.13.9), p.1143, and the 
convolution theorem for Laplace transforms. We obtain

\begin{equation}
\Psi_2(\chi^2) = \frac{1}{2\sqrt{\alpha \beta}} 
\exp \Bigg ( - \frac{\chi^2}{4\alpha \beta} (\alpha + \beta )
\Bigg) I_0 \Bigg ( \frac{\chi^2}{4\alpha \beta} (\beta - \alpha) \Bigg)
\end{equation}

\noindent
where $\alpha$ and $\beta$ are the two eigenvalues of the $\Sigma C$
matrix, and $I_0$ is the zero order $I$ Bessel function.

For this problem, is also easy to obtain the analytic expression 
for the k-order moment of the distribution, using the binomial 
expansion. We obtain

\[
< (\chi^2)^k > = \pi^{-1} \sum_{l=0}^{k} {k \choose l} 
P_1^l P_2^{k-l} \sigma_1^{-2l} \sigma_2^{-2l} \times
\]
\begin{equation}
\times 
\sum_{j=0}^{l} { 2l \choose 2j} (C_{12})^{2l-2j} 2^k |C|^j 
\Gamma (j + 1/2) \Gamma (k-j+1/2) 
\end{equation}

We compare both the distribution function and the $k$-order moments up
to $k=4$ with the values obtained using our approximation.
The results quoted here correspond to the case $P_1 = P_2 = 1$, and 
$\sigma_1^2 = \sigma_2^2 = 1$, so $C_{12}$ varies in the range 
$[ 0, 1 ]$. Nevertheless, the results are completely general.

\begin{figure}
\includegraphics[width=\columnwidth]{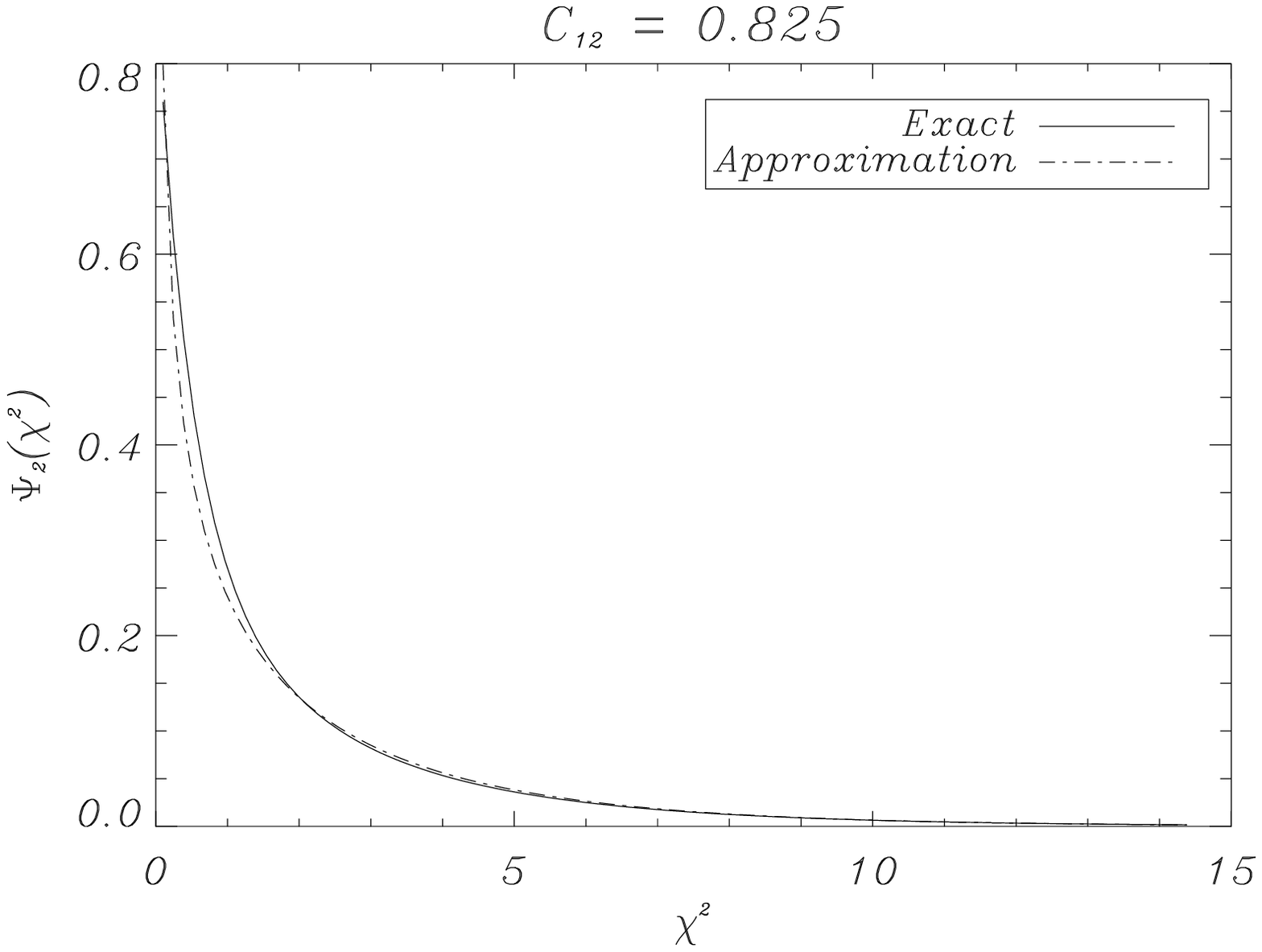}
\caption{Distribution function for a $\chi^2$ with correlations, 
with $N=2$ terms. We have used $P_1 = P_2 = 1$, and 
$\sigma_1 = \sigma_2 = 1$. We show the case $C_{12} = 0.825$,  
because for that value we have the largest percentage difference 
between the true function and our approximation, which is 
given by $n_{eff} = 1.2$ and $A = 0.60$.
We can see that the largest differences occur at low values 
of $\chi^2$. The asymptotic values are well reproduced.} 
\label{app1}
\end{figure}

For our problem, we can write $\alpha = 1 + C_{12}$, 
$\beta = 1 - C_{12}$, and $n_{eff} = 2 ( 1 + C_{12}^2 )^{-1}$.
In Figure \ref{app1} we present the $\Psi_2$ function for the value 
of $C_{12}$ which gives us the maximum percentage difference
between the exact and the approximated functions. This 
value corresponds to $C_{12} = 0.825$.
The largest percentage difference in the 
distribution function for this case is 
reached at $\chi^2 = 0.388$, and has a value of
$\sim 17 \%$. 
In terms of the weights, we 
obtain for this point 
a difference of a $13 \%$ ( $W_{\hat{M}}(true) = 0.73$ , 
and $W_{\hat{M}}(approx) = 0.70$).
Nevertheless, the power of our approximation is that
the largest differences always occur at low values of 
$\chi^2$. The asymptotic shape of the exact 
distribution function is well reproduced, 
as we need for a $\chi^2$ analysis.

\begin{figure*}
\includegraphics[width=\columnwidth]{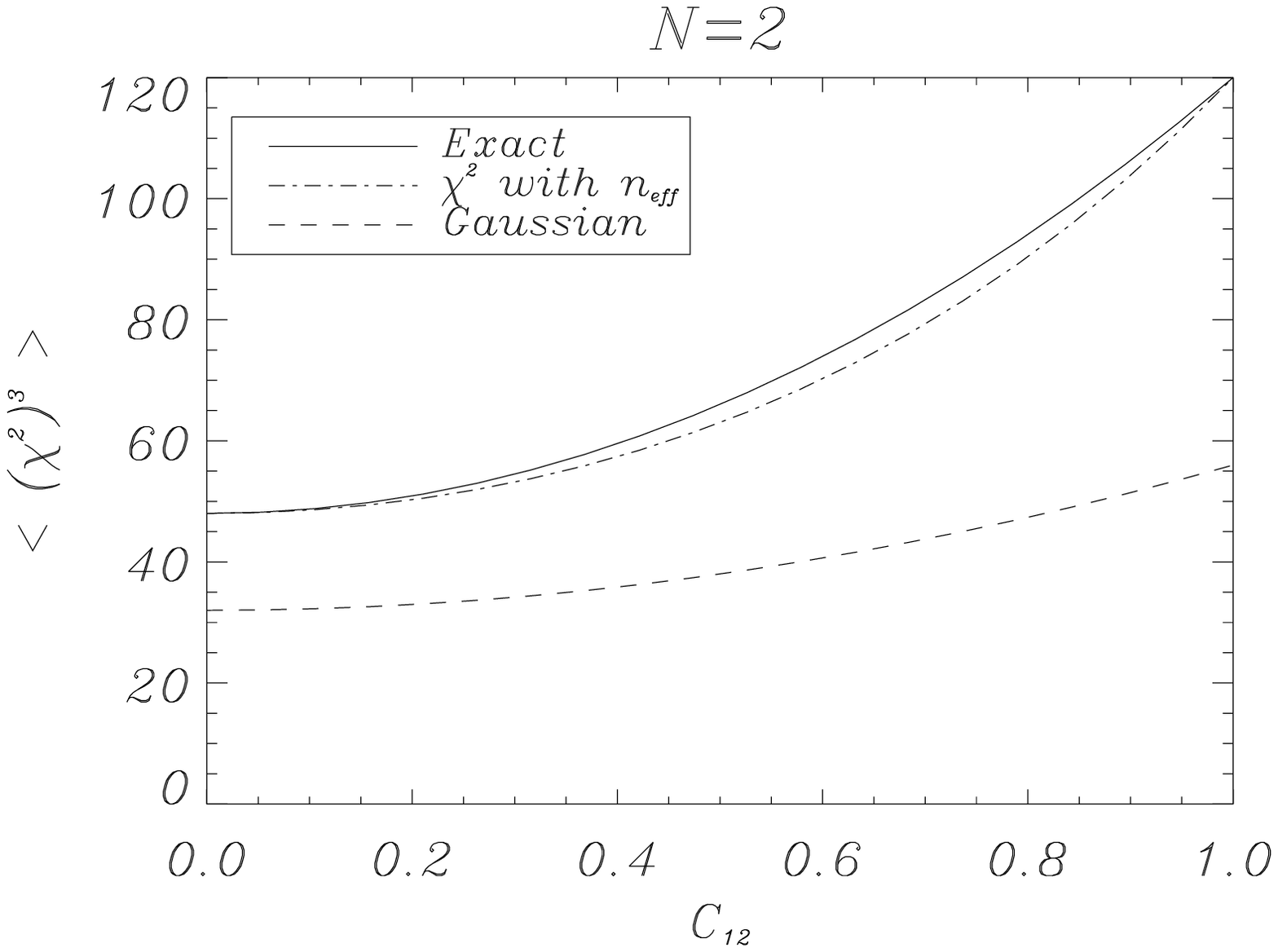}%
\includegraphics[width=\columnwidth]{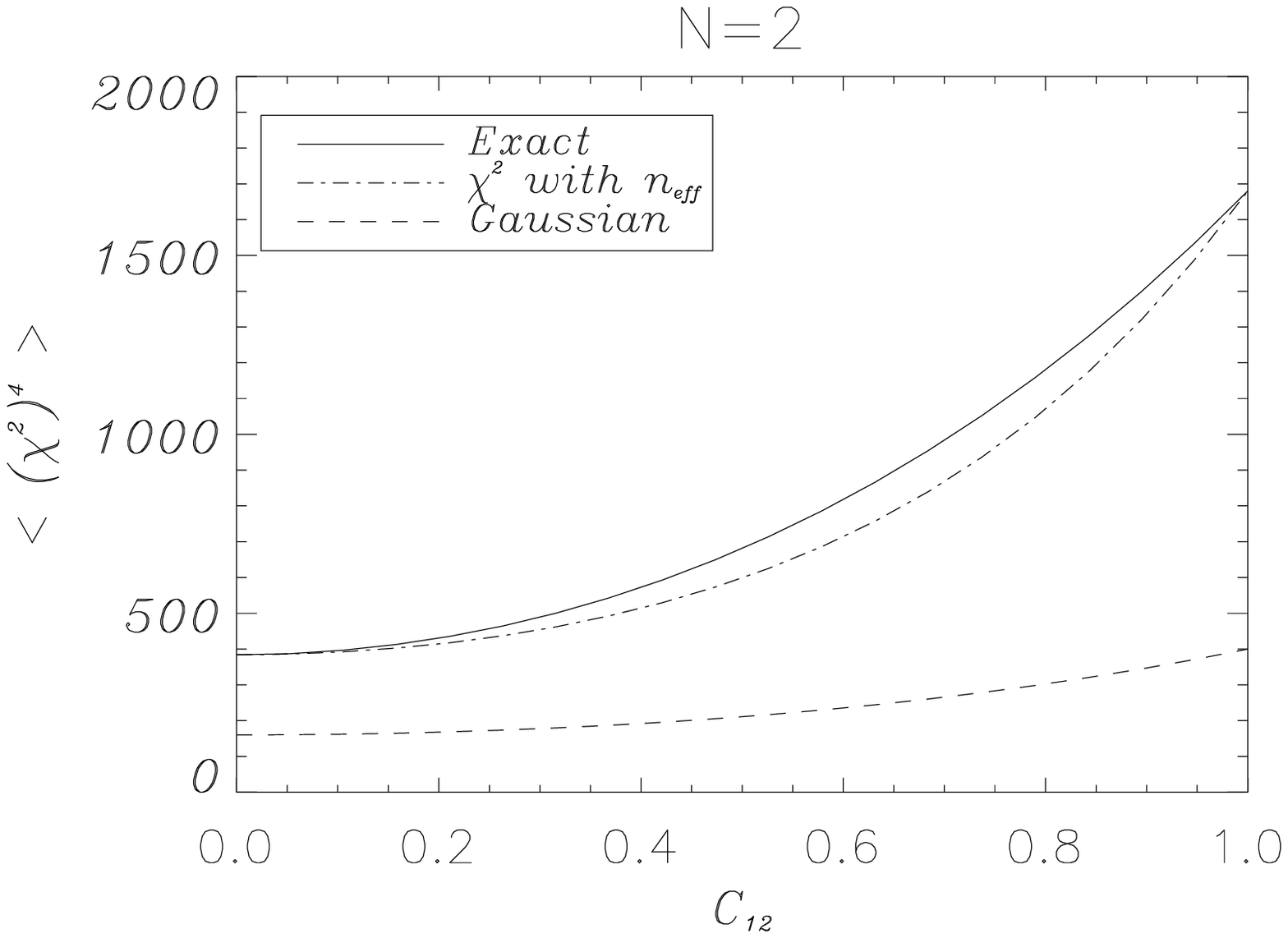}
\caption{Skewness and kurtosis for a $\chi^2$ with $N=2$ terms. 
We see that the approximation and 
the true distribution coincide in the two limit 
cases of $C_{12}=0$ (no correlations) and $C_{12}=1$ (totally 
correlated terms). 
For comparison, we also show these quantities for
a gaussian distribution with the same mean and variance 
as the exact one (see text for details).}
\label{app2}
\end{figure*}

To conclude, we show in Figure \ref{app2} 
the third and fourth order moments, both
for the real distribution and the approximation, in the whole 
range of values for $C_{12}$. By definition, the first and second
moments are equal for the true and the approximate 
distribution. We see again that the 
approximation follows quite closely the 
true function, as we have found from the simulations in Section 5.1.

\end{document}